**Ali R. Baghirzade**
Master of economics
Junior researcher of the research Institute "Innovative Economics" Plekhanov Russian University of Economics Moscow, Russia
e-mail: Bagirzade.AR@rea.ru
**Kushbakov B.**
Master of economics Financial University under the Government of the Russian Federation
e-mail: Beko@gmail.com

# CROWDFUNDING FOR INDEPENDENT PARTIES

**Introduction**

Nowadays there are a lot of creative and innovative ideas of business start-ups or various projects starting from a novel or music album and finishing with some innovative goods or website that makes our life better and easier. Unfortunately, young people often do not have enough financial support to bring their ideas to life. The best way to solve particular problem is to use crowdfunding platforms.

Crowdfunding itself is a way of financing a project by raising money from a "crowd" or simply large number of people.

It is believed that crowdfunding term appeared at the same time as crowdsourcing in 2006. Its author is Jeff Howe. However, the phenomenon of the national funding, of course, much older. For instance, the construction of the Statue of Liberty in New York, for which funds were collected by the people.

Currently, the national project is financed with the use of the Internet. Author of the project in need of funding, can post information about the project on a special website and request sponsorship of the audience. Firstly, author selects the best crowdfunding platform for project requirements and sign in. then he or she creates and draws up the project. The project that is created must correspond to one of the categories available for selection (music, film, publishing, etc.). If you create brand new product, it is necessary to submit the draft-working prototype or sample product. A full list of design rules for a project can be viewed directly on the site of crowdfunding platform. While calculating the cost of project it is necessary to take into account the cost of realization the project, reward for your sponsors, moreover commission of payment systems and taxes. The project is considered successfully launched after it gets through moderation on website.

It is necessary to define the concept of "crowdfunding platform" for considering the mechanisms of work of crowdfunding via Internet. This term refers to a special website that hosts information about projects. This trend has arisen and become widespread in the West. Two largest and most popular crowdfunding platforms are located in the United States: KickStarter (www.kickstarter.com) and Indiegogo (www.indiegogo.com).

*Platform KickStarter*

Kick Starter is a web site for attraction sources for creating various projects using crowdfunding scheme.

Kickstarter helps artists, musicians, filmmakers, designers, and other creators find the resources and support they need to make their ideas a reality. To date, tens of thousands of creative projects — big and small — have come to life with the support of the Kickstarter community.[1]

Kickstarter is an enormous global community built around creativity and creative projects. Over 10 million people, from every continent on earth, have backed a Kickstarter project.

Every artist, filmmaker, designer, developer, and creator on Kickstarter has complete creative control over their work — and the opportunity to share it with a vibrant community of backers.

---

[1] www.kickstarter.com

Since launch, on April 28, 2009, 12 million people have backed a project, $2.7 billion has been pledged, and 113,591 projects have been successfully funded.

Categories represented at this crowdfunding platform: Art, Comics, Crafts, Dance, Design, Fashion, Film & Video, Food, Games, Journalism, Music, Photography, Technology, Theatre. Under the rules of Kickstarter, the project authors retain full rights to their work. The platform can be used to attract financial income, capital or to attract loans.

If the project gets the successful financing, the platform will charge 5% of the amount of donations, in addition to the payment systems, make transfers, charge 3-5% of the transferred funds.

There are three rules that must be followed by each project on KickStarter platform:

1. The project should create something than can be shared with the community.
2. Presentation of the project should be clear and fair.
3. Projects cannot raise money for charity or other financial incentives. As well as cannot contain the prohibited information.

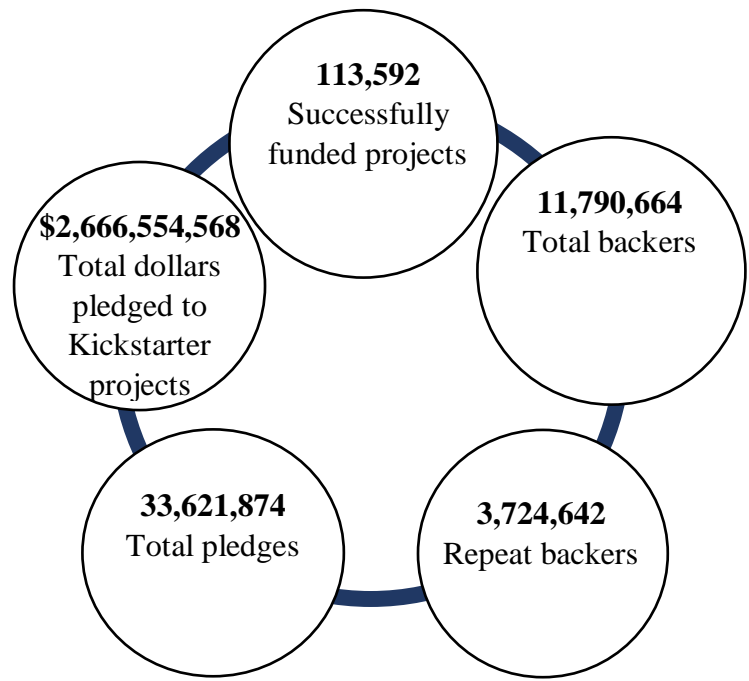

**Picture 1. Kickstarter's statistics**

Source: www.kickstarter.com

*Indiegogo*

Company Indiegogo, as Kicktarter, is one of the largest and the most popular crowdfunding platform. Indiegogo is a platform based on the principles of openness, transparency, freedom of choice and action. The main difference from KickStarter is the absence of restrictions on the type of project, as well as the absence of geographical restrictions on the location of the author. To start the project, you must have a bank account only.

Indiegogo platform was founded in 2008 by a team of three developers: Danae Ringelmann, Eric Schell, Slava Rubin. The basic idea was to allow creative entrepreneurial ideas come to life. The basic idea is contained in the platform slogan: Together, we can do anything!

The platform provides the opportunity to select the payment system: «All or Nothing» or «Flexible Funding». Flexible funding system allows keeping all the contributions made by backers even if the project is not fully funded. The system of tax collection from the author of project is differently calculated depending on the choice of the method of financing. With flexible financing - from 5 to 9%, with an «all or nothing» - 5%.

The project launched on Indiegogo can be of any content. A list of categories is represented by a wide range of areas.

Categories represented on this platform: Tech, Film, Small Business, Community, Music, education, Design, Environment, Gaming, Video / Web and others.

For no-profit projects there is special service Generosity (www.generosity.com) and the commission is not taken for such campaigns.

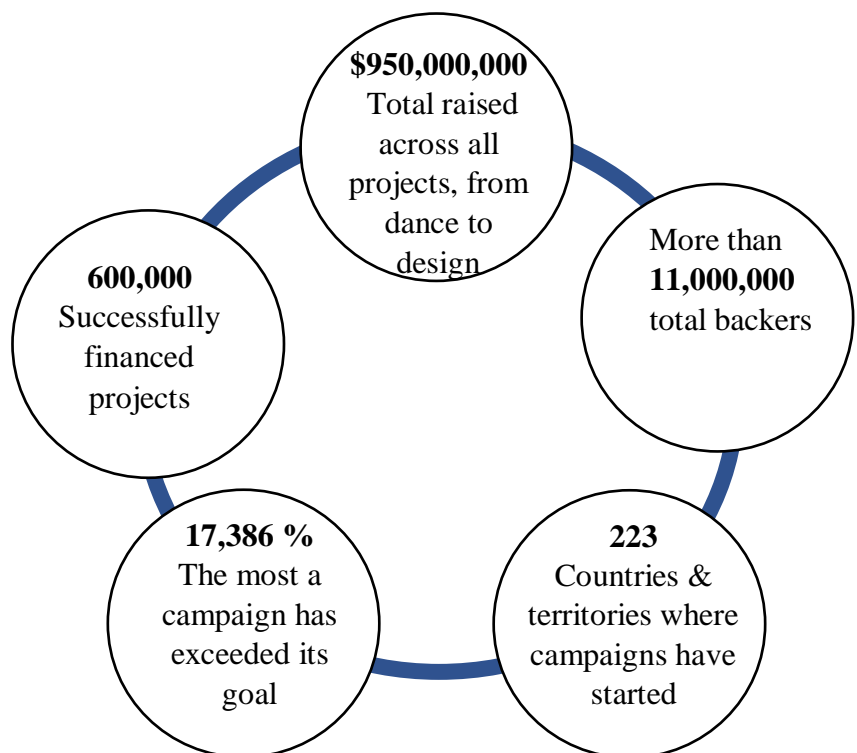

**Picture 2. Indiegogo's statistics**

Source: www.indiegogo.com

In foreign crowdfunding practice, there are exceptional cases of successful projects over the creative teams. Thus, for example, it became widely known case of Amanda Palmer. The joint project of Amanda Palmer and The Grand Theft Orchestra under the name «Theatre Is Evil: the album, art book and tour» gained $ 1,192,793 on KickStarter platform instead of the required amount of $ 100 000.

It was planned to raise funds to record a music album of the singer, an art book with the author's paintings and the tour. During the campaign, the singer was supported by 24.883 sponsor. Amanda Palmer's secret of success defines a fairly simple formula: «stop worrying and let people help you.» There was a creative video on the project's page that was aimed to tell about the campaign and to encourage people to support it. A wide range of sponsorship awards have been developed in various price categories. During the project's campaign constant interaction with the sponsorship audience was carried out on the permanent basis, for example by including additional remuneration limited series. Some rewards include performing a portrait of a sponsor itself singed by Amanda Palmer and its further publication on the project page.

The funds that were raised during the project have been implemented to record album and release the book, which was attended by more than 30 artists, and moreover to conduct the world tour (lasting more than 1 year).

Amanda Palmer in her book «The Art of Asking» tells her personal story of becoming a popular singer, and explains how the ability to solicit and accept leads to success and also establishing trust with her fans.

Crowdfunding in Russia

However, this trend is only gaining popularity in Russia. Founder of Boomstarter crowdfunding platform Eugene Gavrilin, believes that «creative thinking in humans is obtained much more than finding the money to support his ideas».[2] According to him, crowdfunding in Russia in the coming years will completely change the format of interaction between the author and the user of the project.

Nowadays, Russian creative groups wishing to announce the launch of their project, can take advantage of platforms: BoomStarter (www.boomstarter.ru), Planeta ([www.planeta.ru](www.planeta.ru)). They are developed and popular websites. The total amount collected by Planeta.ru platform is 419,999,134 rubles in May 2016.[3] Platforms do not provide detailed information on finance statistic data for public access. Some statistics on the project is available directly from their authors.

Boomstarter

Boomstarter is Russian crowdfunding platform for raising funds in the creative, technical and other kinds of projects; it appears to be an analogue of Kickstarter. The official launch took place on 21 August 2012. Owners of Boomstarter are Ruslan Tugushev and Eugene Gavrilin.

Implemented projects: 1070.

The amount of funds raised: 130 000 000, according to the data of 2015.[4]

Categories represented on the platform are Design, Food, Games, Publishing, Art, Events, Fashion, Music, Society, Sports, Theater, Technology, Movies and video, photography, choreography.

Financial platform philosophy is similar to KickStarter practice: on Boomstarter project is considered successful only when the 100% is financed, the financing model is «all or nothing».

When you run the project on Boomstarter author is able to support an information platform and personal manager-facilitator who helps the author to conduct his project at all stages of implementation.

The rules applicable to the projects Boomstarter platform[5]:

1) a clearly articulated goal (the album recording, edition, creation of works of art).

2) limited timeframe (it is 30 days on average).

3) product prototype (for the project, creating a new product).

There are video lectures revealing the technology of the successful campaign on the platforms in details: it tells how you should design your ad text, page design, video about the project, infographics, development of sponsorship awards; technology promotion project. This material will be useful to the author, who decided to start their campaign on Boomstarter platform[6].

Platform Planeta

Planeta.ru is a social-service platform for collective creation, payment and distribution of digital content and material in Russia. Planeta is one of the first crowdfunding platforms in Russia. The main category of projects are creative projects (music, cinema, theater), moreover, it is possible to carry out charity campaigns on the platform. This platform is an analogue of the international Indiegogo platform[7]. Founders are Fedor Murachkovsky, Max Litmus, Basil Andryushchenko.

Date of launching platform is 7 June 2012.

Number of realized projects are more than 2000, as of April 2016.

Commission of «Planet» and payment aggregators is 10% of the funds collected by the successful project. If the project gathered from 50 to 99% of the whole amount, the total platform fee will be 15%, a flexible funding model projects. On the other hand for charitable projects fee is not charged.

In addition to the actual crowdfunding other services are developed, for example: online-translation online store of unique products, where you can find books and CDs with autographs

---

[2] https://asi.ru/history/9935/#history
[3] https://planeta.ru/
[4] http://www.crowdwillfund.ru/how-to-choose-a-crowdfunding-platform
[5] Crowdfunding LLC, https://boomstarter.ru/help/guidelines
[6] Crowdfunding LLC, https://boomstarter.ru/crowd_learning
[7] РИА Новости, 07.06.2012, http://surfingbird.ru/surf/SWT7e6B2#.V0VY39SLTG

and actions of already completed projects. In addition, the platform is working in the «Charity» category with MegaFon and in the «Social Entrepreneurship» category with Lipton. Also «School crowdfunding Planeta.ru" takes a special place in the work of the platform (planeta.ru/welcome/school.html), after passing this school the author is able to present the idea to the experts in the field of social entrepreneurship and crowdfunding. Successfully presented projects can be supported in promotion by planeta.ru, Fund «Our Future», «Social Information Agency» and «Greenhouses of Social Technologies.»

Despite the relatively young market of crowdfunding, there are already distinguished crowdfunding-authors in Russia. The most striking event in the field of artistic and musical crowdfunding in Russia is a project of Boris Grebenshchikov and group «Aquarium», which collected more than 7 million rubles in 2015 (https://planeta.ru/campaigns/aquarium).

The project was created in order to collect people's money to record a music album and publish the new songs. Initially, the amount requested for the project amounted to 3 000 000 rubles. Boris Grebenshchikov explained that the high cost is related with expensive recording. However, the project quickly exceeded the amount that was requested. A lot of people connect the secret of success with the popularity of Boris Grebenshchikov.

Another example of successful crowdfunding project is a social media Colta.ru - Public Library of artistic, historical, cultural activities, existing since 2013. Each year this social platform receives the funds necessary for the conducting the project through crowdfunding platform Planeta.ru

Crowdfunding opportunities in Russia for independent creative teams very broad. Before starting the project it is necessary to choose the most suitable requirements of crowdfunding platform and get acquainted with the rules of the platform.

Mostly, music and charitable projects have greater success at planeta.ru, and business ideas on Boomstarter[8].

Recommendation for a successful crowdfunding campaign

As a recommendation for beginners I can cite as an example the words of Alexei Dubrovsky, co-founder of the company «CrowdConsulting», head of «Crowdsourcing» projects, advising companies and helping them to make successful projects in the western crowdfunding platforms. He argues that the success of crowd-campaign depends on the correct preparation to it[9].

There is a detailed guide how to make your project successful:[10]

• Research successful and failed campaigns to see which projects, marketing strategies and rewards connected with fans. Analyze why, and to what extent, they work.

• Consider the breakdown of rewards they offer (i.e. physical goods vs. exclusive experiences), cost structure and promotional vehicles used to spread the word about these projects (social vs. traditional media, video endorsements by noted personalities, etc.).

• Budget conservatively up front, including factoring in all costs for reward fulfillment. Leave yourself a 20-30 percent cushion for hidden expenses, then ask for the minimum funds needed to complete your project to make goals seem more attainable to backers.

• Understand who your target audience is, where fans live online and how to reach them. Know what tools you have to get their attention and prepare all supporting assets (videos, screenshots, social media accounts, etc.) in advance.

• Practice and refine your pitch until you can summarize your project in less than 20 seconds. Then build a running promotional campaign that incorporates an ongoing series of marketing activities to run throughout your project's entire duration.

• When it comes to creating killer rewards, use a combination of merchandise, once-in-a-lifetime opportunities and personalized gifts to generate cash and awareness. Offer attention-getting gifts at all reward levels, including impulse-buy levels, and don't leave too many gaps between pricing tiers so everyone has a chance to contribute regardless of personal budget.

---

[8] Nikita Larionov. How to choose a crowdfunding platform? http://www.crowdwillfund.ru/how-to-choose-a-crowdfunding-platform

[9] Denis Boyarinov, Colta.ru [Electronic resource] http://archives.colta.ru/docs/29031

[10] A Beginner's Guide to Crowdfunding, http://www.rollingstone.com/culture/news/a-beginners-guide-to-crowdfunding-20120517

• Keep video and homepage pitches short and sweet. Presentation is everything: Don't skimp on production values. Quickly convey what your project is, why it matters, what qualifies you to make it happen and how it benefits readers/viewers. A picture is worth a thousand words; use video and screens to communicate wherever possible and focus on one to three unique sales points which should be reinforced in all descriptions and promotions.

• Creating a running dialogue with backers, fans and media is vital, as succeeding with crowdfunding requires that you stay at the front of minds. Social media services like Twitter can be even more effective than articles and interviews, as can keeping in constant contact with backers through ongoing updates. A mix of promotional activities should be used to generate chatter on a consistent basis.

This piece of advice gave Amanda Palmer to authors launching their own campaign for the first time[11]:

«... If you're offering a giant array of [reward] options to fans, you have to accept that fulfilling and managing [campaigns] is going to take a large part of your day. You can't phone it in – you have to be realistic about it, and it comes down to running a business. Don't take things for granted».

In an interview in 2013 group BI-2, said: «The only way out is to cooperate with the fans directly»[12]. Having gained the trust of fans, giving them certain bonuses and the opportunity to participate in the creative activity a group wins in the financing of their projects and the sale of their albums through the Internet.

Conclusion

Crowdfunding as a social innovation that helps consumers feel unity with the artist, to feel involvement in the creative process, to support the creative ideas of the author, to help carry out interesting projects gaining more and more popularity in Russia. Fans of creative musicians finance their favorite artists with pleasure.

For the success of the campaign you should follow three basic rules:

Firstly, you need to record a video about the project in which author will appeal to his fans, friends, the community, which author focuses on and ask for help in implementing the project. The video should be concise and dynamic.

Secondly, you need a competent and qualitative description of the project and developed system of sponsorship benefits.

Thirdly, you need to understand not only how to put a project, but also when - at what stage of the project – you should do this.

There must be proper preparation and the right exit time. This is the set of activities that will necessarily lead author's project to success.

**Acknowledgment**

---

[11] www.rollingstone.com

[12] Karpukhina Maria, 3 million in 1.5 days: what is the secret of Boris Grebenshchikov's record, Sobesednik.ru, http://sobesednik.ru/kultura-i-tv/20150904-3-mln-za-15-dnya-v-chem-sekret-rekorda-borisa-grebenshchikov